\documentclass[12pt]{article}

\usepackage[a4paper,margin=2.5cm]{geometry}

\usepackage[T1]{fontenc}
\usepackage{fix-cm}
\usepackage{newtxtext}
\usepackage[amsthm]{newtxmath}

\usepackage{bm}

\usepackage{graphicx}
\usepackage{epstopdf}
\usepackage{caption}
\usepackage{subcaption}
\usepackage[rightcaption]{sidecap}
\usepackage{placeins}

\usepackage{xcolor}
\usepackage{comment}
\usepackage{natbib}
\usepackage{siunitx}
\usepackage{hyperref}

\definecolor{darkgreen}{rgb}{0.0,0.5,0.0}

\hypersetup{
  colorlinks = true,
  urlcolor   = blue,
  citecolor  = black,
}

\newcommand{\grad}{\nabla}
\newcommand{\curl}{\nabla\times}

\newcommand{\vU}{\ensuremath{\mathbf{U}}}

\newcommand{\vnb}{\ensuremath{\mathbf{n}_B}}

\newcommand{\vF}{\ensuremath{\mathbf{F}}}

\newcommand{\vUB}{\ensuremath{\mathbf{U}_B}}
\newcommand{\Uinf}{\ensuremath{U_\infty}}

\newcommand{\Um}{\ensuremath{\overline{\mathbf{U}}}}
\newcommand{\up}{\ensuremath{\mathbf{u}'}}
\newcommand{\Pm}{\ensuremath{\overline{P}}}
\newcommand{\pp}{\ensuremath{p'}}

\newcommand{\wmean}{\ensuremath{\overline{\bm{\omega}}}}
\newcommand{\wvec}{\ensuremath{\bm{\omega}}}

\newcommand{\dS}{\ensuremath{\,\mathrm{d}S}}
\newcommand{\dO}{\ensuremath{\,\mathrm{d}\Omega}}

\newcommand{\vx}{\ensuremath{\mathbf{x}}}

\newcommand{\ek}[1]{\ensuremath{\mathbf{e}_{#1}}}

\newcommand{\Rey}{\ensuremath{\mathit{Re}}}

\newcommand{\phik}{\ensuremath{\phi_k}}

\newcommand{\Fvp}{\ensuremath{F^{(\mathit{vp})}}}
\newcommand{\Fadd}{\ensuremath{F^{(\mathit{add})}}}
\newcommand{\Frs}{\ensuremath{F^{(\mathit{rs})}}}
\newcommand{\Fvisp}{\ensuremath{F^{(\mathit{vis\!-\!p})}}}
\newcommand{\Fvisf}{\ensuremath{F^{(\mathit{vis\!-\!f})}}}
\newcommand{\fvpp}{\ensuremath{f^{(\mathit{vp})}}}
\newcommand{\frsp}{\ensuremath{f^{(\mathit{rs})}}}

\newcommand{\bFvp}{\ensuremath{\mathbf{F}^{(\mathit{vp})}}}
\newcommand{\bFrs}{\ensuremath{\mathbf{F}^{(\mathit{rs})}}}
\newcommand{\bFvisp}{\ensuremath{\mathbf{F}^{(\mathit{vis\!-\!p})}}}
\newcommand{\bFvisf}{\ensuremath{\mathbf{F}^{(\mathit{vis\!-\!f})}}}

\newcommand{\fvpy}{\ensuremath{\Lambda_{L,y}\,\omega_y}}
\newcommand{\frsz}{\ensuremath{\phi_{L,z}\,\zeta_z}}

\newcommand{\Bnabla}{\ensuremath{\boldsymbol{\nabla}}}
\newcommand{\Bdiv}{\ensuremath{\Bnabla\!\cdot\,}}
\newcommand{\Bgrad}{\ensuremath{\Bnabla\,}}
\newcommand{\Bcurl}{\ensuremath{\Bnabla\!\times\,}}

\newcommand{\taurs}{\ensuremath{\bm{\tau}}}
\newcommand{\zRS}{\ensuremath{\bm{\zeta}}}

\title{The Reynolds-Averaged Vortex Force Map Method}

\author{
Matteo Liguori, Zhan Zhang, Francesco Ciriello, Juan Li\thanks{Corresponding author: \texttt{juan.li@kcl.ac.uk}}\\
Department of Engineering, King's College London
}

\date{}

\begin{document}
\maketitle
\begin{abstract}

Vortex-force mapping (VFM) links vortical flow structures to aerodynamic forces through compact-domain integrals weighted by geometry-only Laplace potentials, but existing formulations are tied to simple geometries and laminar flows.  
In this study, we derive a Reynolds-averaged vortex force map (RA-VFM) directly from the incompressible Reynolds-averaged Navier-Stokes (RANS) equations, augmenting the classical vortex-pressure (VP) term with a Reynolds-stress (RS) contribution based on the Laplace-potential-weighted divergence of the modelled Reynolds stress (Boussinesq eddy-viscosity form). The resulting framework reconstructs mean lift and drag from RANS mean fields while retaining spatial attribution of force production to specific regions and coherent structures within a compact control volume. 
We apply RA-VFM to unsteady RANS ($k$-$\omega$ SST) simulations of a realistic gliding goshawk with strong three-dimensionality and a matched GOE803 aerofoil section.
 For the aerofoil, the VP term alone reproduces the CFD force curves over the pre- and near-stall range, with RS contributions becoming appreciable only in deep stall. 
 For the bird, by contrast, the VP term underpredicts both $C_L$ and $C_D$, whereas including the RS term reduces the mean absolute error relative to CFD from $6\%$ to $2\%$ in lift and from $5\%$ to $1\%$ in drag over an angle of attack range of $0^\circ$-$20^\circ$. RA-VFM thus extends vortex-force mapping to turbulent, 3-D RANS flows and enables quantitative attribution of mean lift and drag to specific coherent structures within compact domains.
\end{abstract}

\newpage
\section{Introduction}
The extraction of body-forces from flow field data is crucial in fluid dynamics, with various applications in aerodynamic design, flow control, and bio-inspired engineering. Various methods exist for force extraction, including integration of pressure and skin friction, impulse methods, and momentum balance approaches \citep{Luckring2019, Chen2023}. The integration of surface pressure and skin friction is effective, especially in computational fluid dynamics (CFD); but requires pressure field information, which can be challenging to obtain due to the sensitivity of the pressure Poisson equation to spatial resolution and boundary conditions \citep{Fujisawa_2005}. The classical momentum balance method requires time-resolved velocity data and poses difficulties with boundary conditions and spatial resolution, yet it has been successfully applied in some studies \citep{UNAL199,vanOudheusden2007}. The impulse method calculates forces from the vorticity field without iterative pressure computations, providing accurate estimates \citep{NOCA1997345}, but requires time-resolved flow field data and full vorticity capture, limiting its applicability to starting flows. A family of volume-based integral force diagnostics method has been explored \citep{GaoXieLu2025WeightedIntegral,GaoWu2019Galilean}.  

To accurately extract force from velocity and vorticity flow field data, a vortex force map (VFM) method has been proposed by \cite{Li2015, Li2016}. The method is capable of linking vortex structures to their body-force contributions in laminar flows or simple geometries \citep{Li2015,Li2016,Li2018,Li2020b, Li2020a}. It calculates the body-force as an integral of the scalar product of a flow-independent vector and the local velocity multiplied by the vorticity within a finite domain enclosing the body. The method offers rapid force prediction from only a compact domain around the object, and is effective on sparse and noisy data such as particle image velocimetry (PIV) \citep{Otomo2024}. Initially applied to inclined flat plates and simple aerofoils in two-dimensional (2-D) inviscid flows \citep{Li2015, Li2016}, it was later refined for viscous flows and general aerofoils, incorporating \cite{HOWE1995} integral force formula to ensure rapid decay of vortex influence with distance \citep{Li2018}. 
The method showed good agreement with CFD predictions, even in separated flows. Extensions include calculating vortex contributions to moments \citep{Li2020a}, application to three-dimensional (3-D) flows \citep{Li2020b}, and adaptations for multi-body assemblies \citep{Li2022}. Despite these developments, the method has not yet been thoroughly tested on high Reynolds number turbulent flows with complex geometries and under predicts lift and drag when applied to RANS mean fields.

In this work, we derive a Reynolds-averaged vortex force map (RA-VFM) by extending the classical VFM with an explicit Reynolds-stress (RS) contribution that augments the vortex-pressure (VP) and viscous terms (\textsection~\ref{sec:VFM/RAVFM}). We apply RA-VFM to turbulent flow about a gliding goshawk (\textit{Accipiter gentilis}) and to an incidence-matched 2-D GOE803 section representative of the mid-span profile at $\Rey=\mathcal{O}(10^{5})$. Unsteady RANS (URANS) simulations provide the mean fields for assessment and are validated against available measurements and reference data (\textsection~\ref{sec:CFD}); prior studies indicate that higher-fidelity approaches (e.g.\ LES) add little for this configuration \citep{Usherwood2020,Cheney2021}. Force reconstruction accuracy and vortex-structure attribution are reported in \textsection~\ref{sec:results_discussion}.

\section{Methodology}
\label{sec:methodology}
\begin{figure}
\centering
\includegraphics[width=1\linewidth]{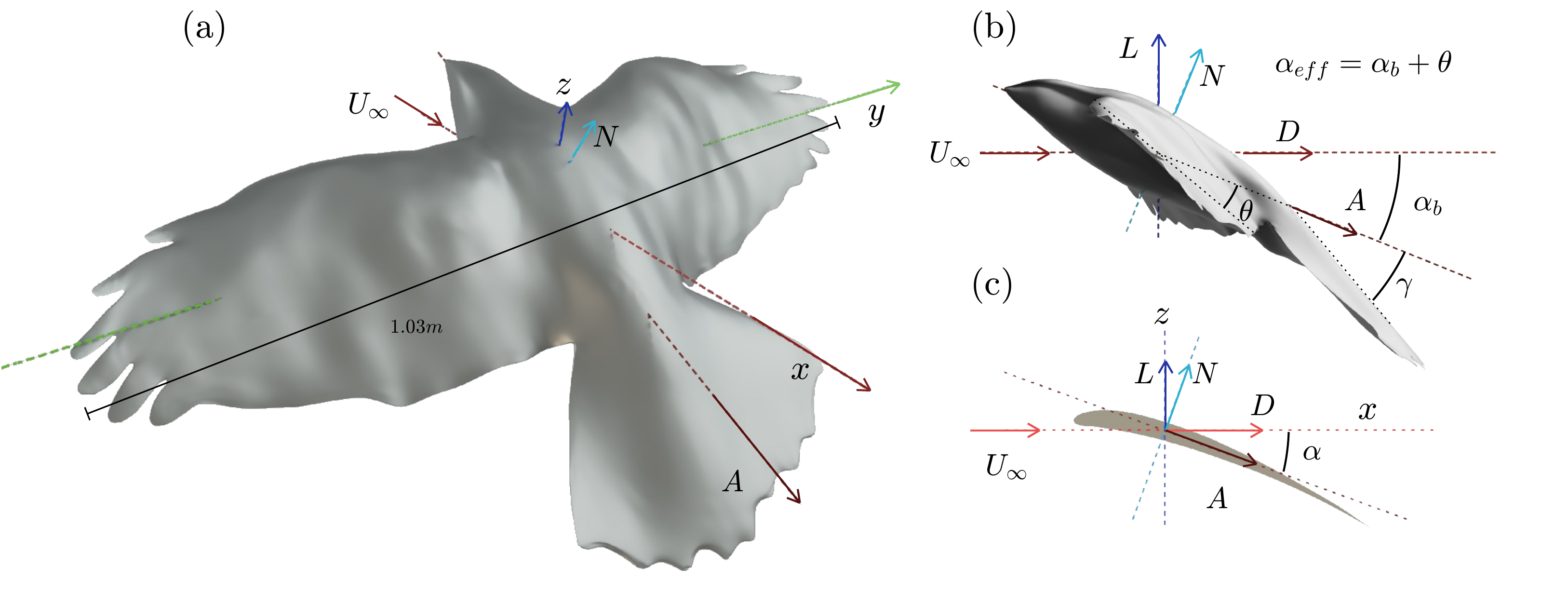}
\caption{(a) 3-D schematic of the \textit{Accipiter gentilis} (Northern Goshawk) in gliding configuration. The ground-fixed coordinate system is defined with $x$ along the freestream direction ($\Uinf$), 
$y$ along the spanwise axis, and $z$ perpendicular to both.
(b) Decomposition of the aerodynamic force into lift ${L}$ and drag ${D}$, 
or equivalently into normal ${N}$ and axial ${A}$ components. The body angle of attack is $\alpha_b$, defined between $\Uinf$ and the body $x$-axis; $\theta$ is the wing–body setting angle and $\gamma$ the body–tail setting angle. (c) Cross-sectional view of the wing, represented by a GOE803 aerofoil profile.}
\label{fig:Geometry}
\end{figure}

We consider incompressible flow around the \textit{Accipiter gentilis} (Northern Goshawk) in a gliding configuration (Figure \ref{fig:Geometry}) with a constant fluid density $\rho=\SI{1.058}{kg\,m^{-3}}$. The velocity of the flow in the ground-fixed frame is
$\bm{U}= (u,v,w)$ and the vorticity is $ \bm{\omega} = (\omega_x ~,~\omega_y ~,~\omega_z) = (w_{,y}- v_{,z}~,~ u_{,z}- w_{,x}~,~ v_{,x}- u_{,y})$.
\subsection{The RA-VFM method}
\label{sec:VFM/RAVFM}
The original VFM method derived in \cite{Li2020b} expresses the aerodynamic force in the $k$-direction ($\mathbf{k}$) as the sum of four components: an added-mass term $\Fadd_k$ (arising from body acceleration), a VP term $\Fvp_k$ (associated with the surrounding vortex structures), a viscous pressure term $\Fvisp_k$, and a skin-friction term $\Fvisf_k$ (generated by viscous effects), each term involving an integration over the body surface $S_B$ or the fluid domain $\Omega$ (with $\dS$ denoting the surface-area element on $S_B$ and $\dO$ the volume element):
\begin{align}
F_k = \Fadd_k + \Fvp_k + \Fvisp_k + \Fvisf_k ,
\end{align}
\begin{subequations}
\begin{align}
\Fadd_k &= \iint_{S_B} \phik \,\frac{\partial \vUB}{\partial t} \cdot \vnb \,\dS, \label{eq:AM}\\
\Fvp_k  &= \rho \iiint_{\Omega} \Bgrad \phik \cdot (\wvec \times \vU) \,\dO, \label{eq:VF}\\
\Fvisp_k &= \mu \iint_{S_B} \phik \,(\Bcurl \wvec)\cdot \vnb \,\dS, \label{eq:Visc1}\\
\Fvisf_k &= \mu \iint_{S_B} (\vnb \times \wvec)\cdot \ek{k} \,\dS. \label{eq:Visc2}
\end{align}
\end{subequations}
Here, $\phi_k$ represents the hypothetical potential obtained by numerically solving the Laplace equations, $\Bnabla^2 \phi_k = 0$, subject to boundary conditions that define the velocity potential of the body moving in the $k$-direction with unit velocity \citep{HOWE1995}. By definition, $\phi_k$ is solely dependent on the geometry and remains independent of the actual flow conditions. 

Here we derive the RA-VFM from the RANS equations. In turbulent flows, the instantaneous fields are decomposed into mean ($\Um$) and fluctuating ($\up$) parts such that, 
\begin{gather}\label{eq:RAVFM_decomp}
\vU(\vx,t) =
\begin{bmatrix}
u(\vx,t)\\[0.5ex]
v(\vx,t)\\[0.5ex]
w(\vx,t)
\end{bmatrix}
=
\begin{bmatrix}
\overline{U}(\vx) + u'(\vx,t)\\[0.5ex]
\overline{V}(\vx) + v'(\vx,t)\\[0.5ex]
\overline{W}(\vx) + w'(\vx,t)
\end{bmatrix}
= \Um(\vx) + \up(\vx,t), \\[1.5ex]
P(\vx,t) = \Pm(\vx) + \pp(\vx,t),
\end{gather}
where the velocity field satisfies the incompressible continuity equation:
\begin{equation}\label{eq:RAVFM_incomp}
\Bnabla\!\cdot\!\Um = 0, \qquad \Bnabla\!\cdot\!\up = 0.
\end{equation}
The instantaneous hydrodynamic force on the body is written as the sum of the instantaneous pressure and skin-friction contributions:
\begin{equation}\label{eq:RAVFM_force_surface}
\vF
= \iint_{S_B} P\,\vnb\,\dS
+ \mu \iint_{S_B} \bigl(\vnb \times \wvec\bigr)\,\dS,
\end{equation}
where $\vnb$ denotes the outward normal vector. To recast the pressure contribution in \eqref{eq:RAVFM_force_surface} into a vorticity-based form, we use the Lamb-Gromyko representation of the incompressible RANS momentum equation:
\begin{equation}\label{eq:RANS_mean}
\grad\!\Bigl(\Pm + \tfrac12\rho\,\Um^2\Bigr)
+ \rho\,\bigl(\wmean \times \Um\bigr)
= -\,\rho\,\Bdiv\overline{\up\up}
\;-\; \mu\,\curl\wmean.
\end{equation}
Similarly, we introduce the hypothetical potentials $\phi_{k}$ which satisfy Laplace's equation,
$\Bnabla^2 \phik = 0,$ with the boundary conditions:
\begin{equation}\label{eq:phik_bc}
\left\{
\begin{aligned}
-\,\Bnabla \phi_k \cdot \mathbf{n}_B 
&= e^k \cdot \mathbf{n}_B = n_{B,k}, 
&\quad& \text{ } S_B,\\
\phi_k 
&\to 0, 
&\quad& \text{} |\mathbf{x}| \to \infty.
\end{aligned}
\right.
\end{equation}
Taking the inner product of~\eqref{eq:RANS_mean} and $\grad\phi_{k}$, and integrating over the fluid domain $\Omega$ gives
\begin{gather}
\iiint_{\Omega} \grad\phik \!\cdot\! \grad\Bigl(\Pm + \tfrac12 \rho\,\Um^2\Bigr)\,\dO
+ \rho \iiint_{\Omega} \grad\phik \!\cdot\! (\wmean \times \Um)\,\dO \notag\\
=\; -\,\rho \iiint_{\Omega} \grad\phik \!\cdot\! \Bdiv\overline{\up\up}\,\dO
\;-\; \mu \iiint_{\Omega} \grad\phik \!\cdot\! (\curl \wmean)\,\dO.
\label{eq:int_phi_RANS}
\end{gather}
Applying vector identities and the divergence theorem, the pressure term is expressed as
\begin{gather}
\iint_{S_B} \Pm\,n_{B,k}\,\dS
= \rho \iiint_{\Omega} \grad\phik \!\cdot\! (\Um \times \wmean)\,\dO + \rho \iiint_{\Omega} \grad\phik \!\cdot\! \Bdiv\overline{\up\up}\,\dO \notag\\
\qquad + \mu \iint_{S_B} \Bigl[(\vnb \times \wmean)\!\cdot\!\grad\phik\Bigr]\,\dS.
\label{eq:surf_integral}
\end{gather}
Substituting~\eqref{eq:surf_integral} into the force definition
\eqref{eq:RAVFM_force_surface} yields the RA-VFM decomposition
\begin{align}\label{eq:RAVFM_vector}
F_k
= \Fvp_{k} + \Frs_{k} + \Fvisp_{k} + \Fvisf_{k} .
\end{align}
\begin{subequations}\label{eq:RAVFM_parts}
\begin{align}
&\Fvp_{k} =
 \rho \iiint_{\Omega} \grad\phik \!\cdot\! (\Um \times \wmean)\,\dO,
 \label{eq:RAVFM_vp}\\[3pt]
&\Frs_{k} =
 \iiint_{\Omega} \grad\phik \!\cdot\! \zRS \,\dO,
 \label{eq:RAVFM_rs}\\[3pt]
&\Fvisp_{k} =
 \mu \iint_{S_B} \Bigl[(n_{Bk} \times \wmean)\!\cdot\!\grad\phik\Bigr]\,\dS,
 \label{eq:RAVFM_visp}\\[3pt]
&\Fvisf_{k} =
 \mu \iint_{S_B} (n_{Bk} \times \wmean)\,\dS.
 \label{eq:RAVFM_visf}
\end{align}
\end{subequations}
Here $\taurs= - \rho\overline{\up\up}\;=\; 2\mu_t \mathbf{S}
\;-\; \tfrac{2}{3}\rho k\,\mathbf{I}$ is the Reynolds-stress under Boussinesq eddy-viscosity approximation; its divergence is expressed as, 
\begin{equation}\label{eq:zeta_def}
\zRS \ :=\;-\rho\,\Bdiv\overline{\up\up} =\Bdiv\taurs,
\end{equation}
where the mean rate-of-strain tensor 
\begin{equation}\label{eq:Sij_def}
\mathbf{S}
\;=\;
\tfrac12\Bigl(\Bnabla\Um + (\Bnabla\Um)^{T}\Bigr)
\; i.e.\;
S_{ij}
\;=\;
\tfrac12\bigl(\partial_j U_i + \partial_i U_j\bigr).
\end{equation}
$\bFvp$ is the VP contribution and $\bFrs$ is the Boussinesq form of the RS contribution. $\bFvisp$ and $\bFvisf$ account for viscous pressure and friction effects and are very small at higher $\Rey$ numbers. This decomposition mirrors the VFM structure while extending it to the Reynolds-averaged (turbulent) setting.

\subsection{CFD and its validation}
\label{sec:CFD}
The CFD analyses for both the 2-D aerofoil and the bird were conducted in STAR-CCM+ using Unsteady Reynolds-averaged Navier-Stokes (URANS) with the $k$–$\omega$ shear–stress transport (SST) model and a Pressure-Implicit with Splitting of Operators (PISO) algorithm, with time step $\Delta t=5\times10^{-5}\,$s.
Mesh independence was checked, leading to a final choice of 0.70 million cells for the aerofoil and 5.4 million cells for the bird. Prism layers were applied near the surfaces to ensure accurate wall treatment, achieving $y^+<0.2$ and Courant number below 3. Both simulations were run at a gliding speed of $7.5~\mathrm{m/s}$ over effective angles of attack from $0^\circ$ to $22^\circ$.
The Laplace equations in the lift and drag directions are solved separately to obtain the hypothetical potentials $\phi_L$ and $\phi_D$.
Lift and drag coefficients are defined as
\begin{equation}
C_L = \frac{L}{\tfrac{1}{2}\,\rho \,\Uinf^2 S}, \qquad
C_D = \frac{D}{\tfrac{1}{2}\,\rho \,\Uinf^2 S}.
\end{equation}
\begin{figure}
\centering
\includegraphics[width=0.8\linewidth,]{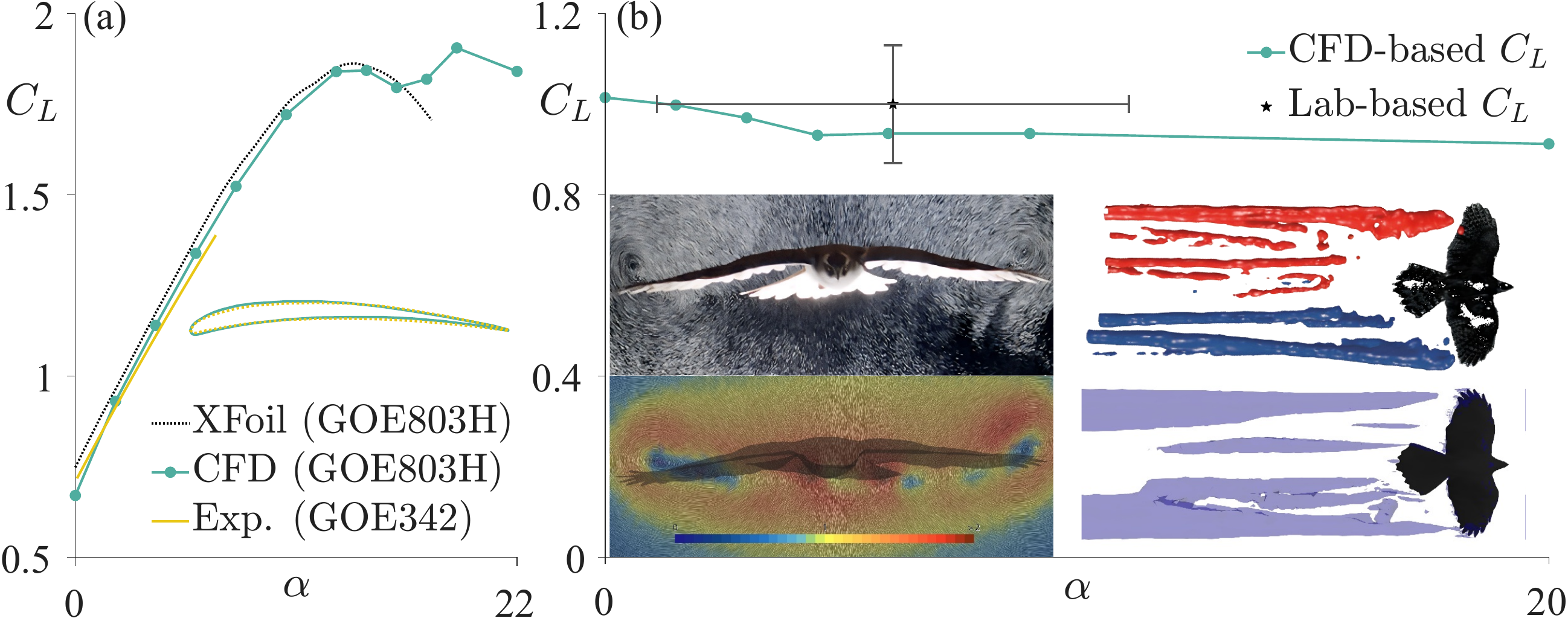} 
\caption{CFD validation against theoretical and experimental results~\citep{Riegels1961,Usherwood2020,Cheney2021} (a) Aerofoil, (b) Bird with wake comparison: CFD (top left), lab measurements (bottom left), and Q-criterion comparison: CFD (bottom right), PTV (top right).}
\label{fig:CFD_Val}
\end{figure}

In our gliding case $F_k^{(add)} = 0$. The viscous effects can be ignored at $\Rey = 2.7\times 10^5$.
The CFD results have been validated 
against experimental data, for the 2-D aerofoil \citep{Riegels1961} and for the bird \citep{Usherwood2020,Cheney2021} in Figure~\ref{fig:CFD_Val}. 
The CFD results for the lift coefficient closely matched experimental lift coefficients for both cases. 
Additionally, the 3-D flow structures and downwash patterns in the CFD align with the particle tracking velocimetry (PTV) data, with vortex structures identified using the Q-criterion (Figure~\ref{fig:CFD_Val}(b)).
\section{Results and Discussion}
\label{sec:results_discussion}

\begin{figure}
\centering
\includegraphics[width=1\linewidth]{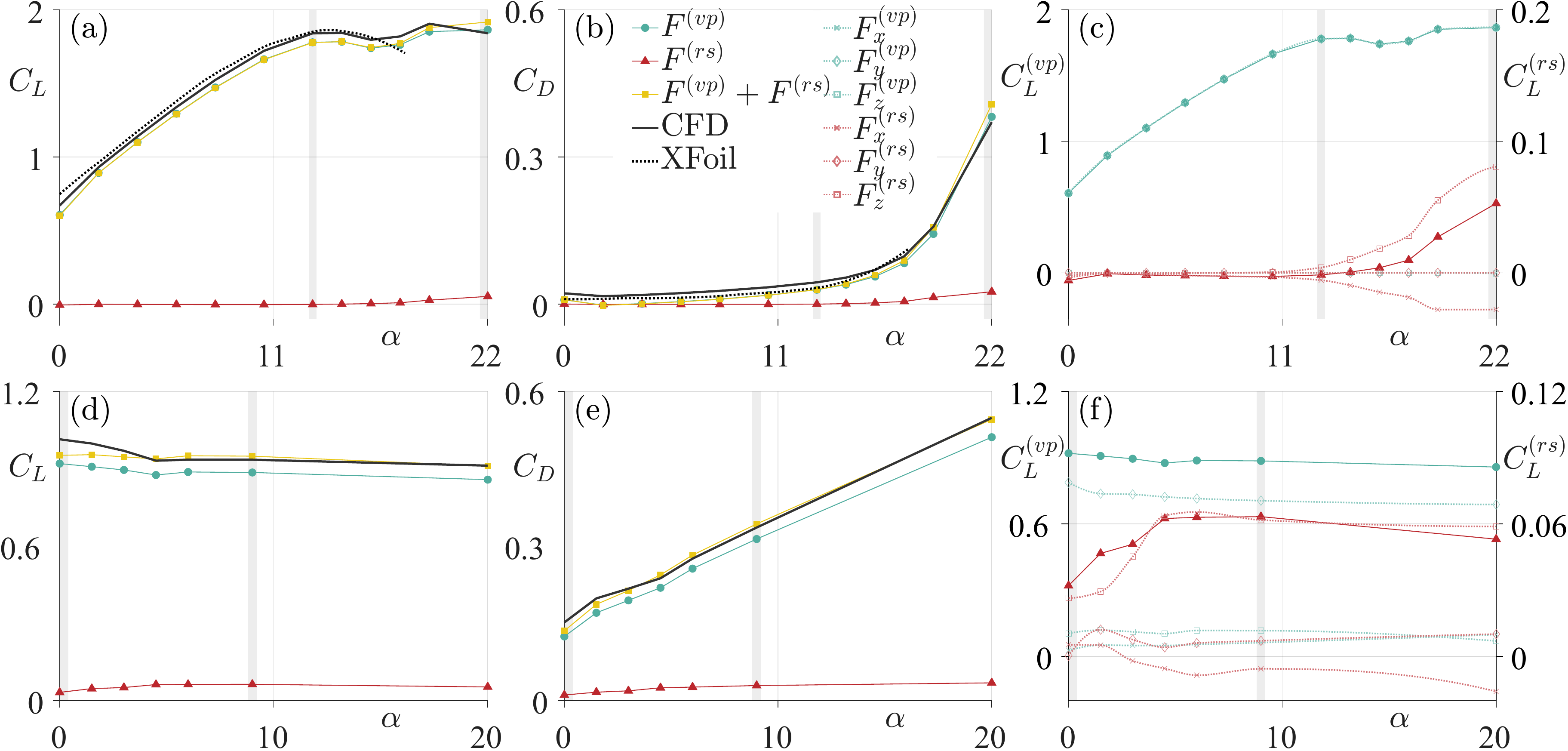}
\caption{Comparison of force coefficients from CFD and RA-VFM method ($\Fvp_k + \Frs_k$) for both the aerofoil (a–c) and the bird (d–f). (a,c,d,f) $k=L$; (b,e) $k=D$. For the aerofoil, XFOIL predictions are included. (c, f) Lift-component breakdown in the $(x, y, z)$ directions, where dual vertical axes are used to represent $\Fvp_L$ and $\Frs_L$ separately.}
\label{fig:Results_plots}
\end{figure}

\begin{figure}
\centering
\includegraphics[width=1\linewidth]{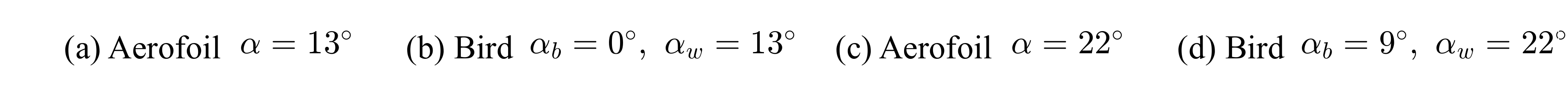}
\includegraphics[width=1\linewidth]{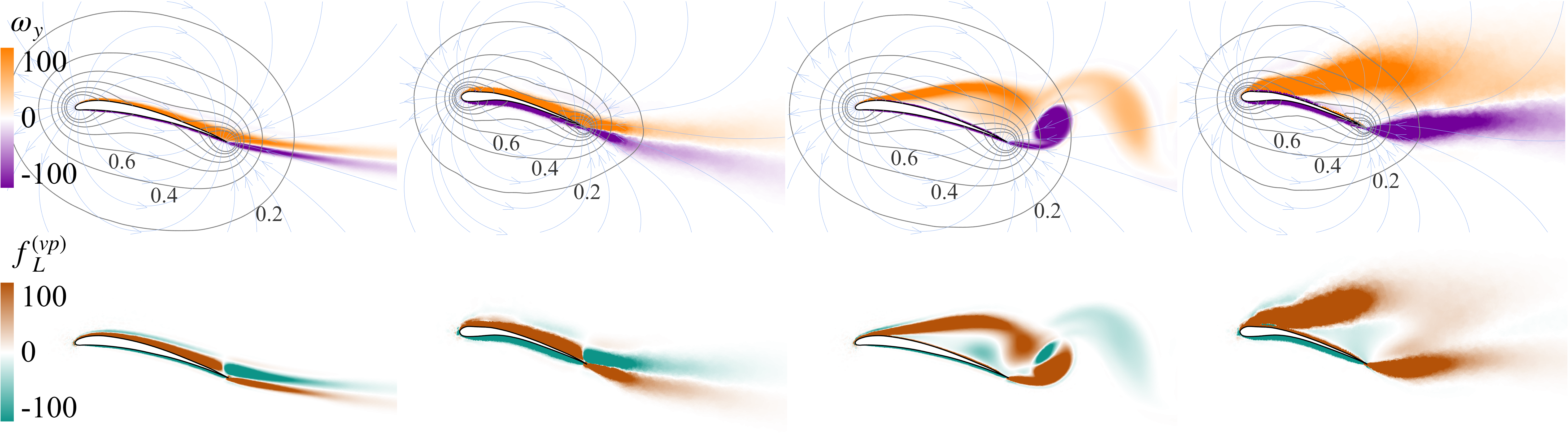}

\vspace{0.3 cm}

\includegraphics[width=1\linewidth]{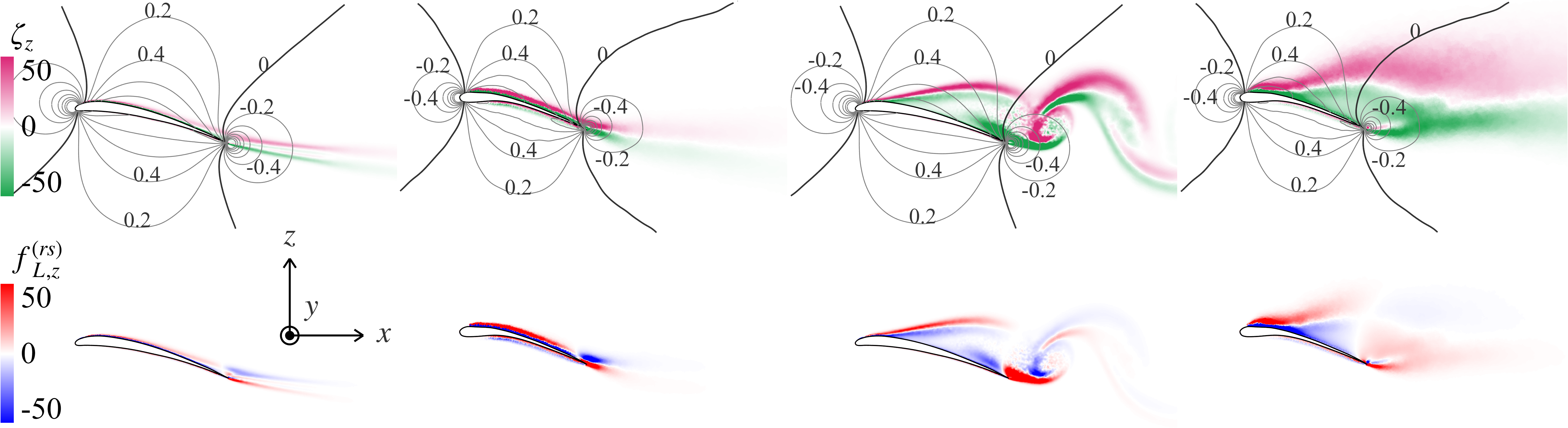}
 \caption{Visualisation of flow fields and dominant lift components for the aerofoil (a,c) and bird (b,d) under incidence-matched conditions ($\alpha_{\mathrm{eff}} = \alpha = 13^{\circ}$ and $22^{\circ}$). 
$1^{st}$ row: contours of $\omega_y$ overlaid with VFM for lift, contour lines show $|\Lambda_{L,y}|$ and arrows show its direction. 
$2^{nd}$ row: distribution of vortex–pressure lift density $f^{(\mathrm{vp})}_L$, mainly contributed by $f^{(\mathrm{vp})}_{L~y}$. 
$3^{rd}$ row: contours of $\zeta_z$ overlaid with contour lines of the gradient of the potential field $\phi_{L,z}$. 
$4^{th}$ row: Reynolds–stress lift density $f^{(\mathrm{rs})}_{L~z} = \phi_{L,z}\,\zeta_z$.
} 
 \label{fig:Results}
\end{figure}

The RA-VFM predictions for both lift and drag are shown in Figure~\ref{fig:Results_plots}, (a,d) and (b,e).

RA-VFM augments VFM with a RS correction and predicts well against CFD and theoretical force coefficients in both cases. For the aerofoil, the $\fvpp$ alone reproduces the CFD curves across the tested range of $\alpha$, while $\frsp$ is negligible except in deep stall ($\alpha>20^\circ$).
For the bird, $\fvpp$ alone underpredicts both $C_L$ and $C_D$, whereas RA-VFM with the $\frsp$ correction matches CFD better for $0^\circ\!\le\!\alpha_b\!\le\!20^\circ$; mean absolute error relative to CFD decreases from $6\%$ to $2\%$ for lift and from $5\%$ to $1\%$ for drag by including the correction.

Despite the strong geometric similarity between the wing section of the goshawk and the aerofoil, their $C_L(\alpha)$ and $C_D(\alpha)$ trends differ markedly due to 3-D effects. 
The bird exhibits a stable $C_L$ and a nearly linear increase in $C_D$, whereas the aerofoil shows typical 2-D characteristics: a linear $C_L$ at small $\alpha$ that saturates near stall, and an upward-curving $C_D$.

The component plots in Figure~\ref{fig:Results_plots} (c,f) show the component-wise decomposition of $\Fvp_k=\iiint_{\Omega}\fvpp_k\dO$ for $k=L$, where $\fvpp_k$ is the VP force density.

\begin{equation}
\fvpp_{k}
= -\rho
\left(
\Lambda_{k,1}\!\cdot\!(0,\,\overline{V},\,\overline{W})\,\omega_x 
+~~ \Lambda_{k,2}\!\cdot\!(\overline{U},\,0,\,\overline{W})\,\omega_y
+~~ \Lambda_{k,3}\!\cdot\!(\overline{U},\,\overline{V},\,0)\,\omega_z
\right),
\label{eq:vpL_component}
\end{equation}
and decomposition of the term $\Frs_{k} = \iiint_{\Omega} \frsp_{k} \, d\Omega$ in~\ref{eq:RAVFM_rs}
\begin{equation}
\begin{aligned}
\frsp_{k} &=
\begin{bmatrix}\phi_{k,x}\\ \phi_{k,y}\\ \phi_{k,z}\end{bmatrix}\!\cdot\!
\begin{bmatrix}
\partial_x\!\big(2\mu_t \,\overline{U}_{,x}-\tfrac{2}{3}\rho k\big)
+\partial_y\!\big(\mu_t(\overline{U}_{,y}+\overline{V}_{,x})\big)
+\partial_z\!\big(\mu_t(\overline{U}_{,z}+\overline{W}_{,x})\big)\\[3pt]
\partial_x\!\big(\mu_t(\overline{U}_{,y}+\overline{V}_{,x})\big)
+\partial_y\!\big(2\mu_t \,\overline{V}_{,y}-\tfrac{2}{3}\rho k\big)
+\partial_z\!\big(\mu_t(\overline{V}_{,z}+\overline{W}_{,y})\big)\\[3pt]
\partial_x\!\big(\mu_t(\overline{U}_{,z}+\overline{W}_{,x})\big)
+\partial_y\!\big(\mu_t(\overline{V}_{,z}+\overline{W}_{,y})\big)
+\partial_z\!\big(2\mu_t \,\overline{W}_{,z}-\tfrac{2}{3}\rho k\big)
\end{bmatrix},
\end{aligned}
\label{eq:zeta-expanded}
\end{equation}
for the aerofoil and the bird wing section, respectively. The VP lift contribution consists of three components defined by the vorticity in $x,y,z$ directions and the corresponding Lambda vector defined in~\ref{eq:app_vp_Lambda}. For the aerofoil, the VP lift contributions from $\omega_x$ and $\omega_z$ are $0$, as it is quasi-2-D.
The RS lift contribution consists of three components, obtained by multiplying the divergence of the RS tensor ($\zeta$) in the $x,y,z$ directions with the corresponding components of the gradient of $\phi_L$.
For both aerofoil and bird, the contribution from $\zeta_z$ is dominant, that from $\zeta_y$ is negligible, while that from $\zeta_x$ is weaker and of opposite sign to $\zeta_z$. 
This force decomposition helps diagnose the origin of forces in the flow by mapping the VP and RS densities onto the potential gradient field. The dominant VP and RS lift sources, $\fvpy$ and $\frsz$, are examined in greater detail in Figure~\ref{fig:Results}.

\begin{figure}
\centering
\includegraphics[width=1\linewidth]{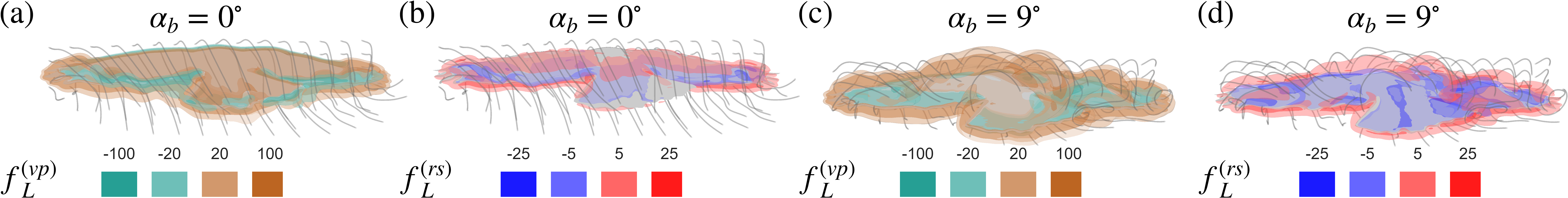}
\includegraphics[width=1\linewidth]{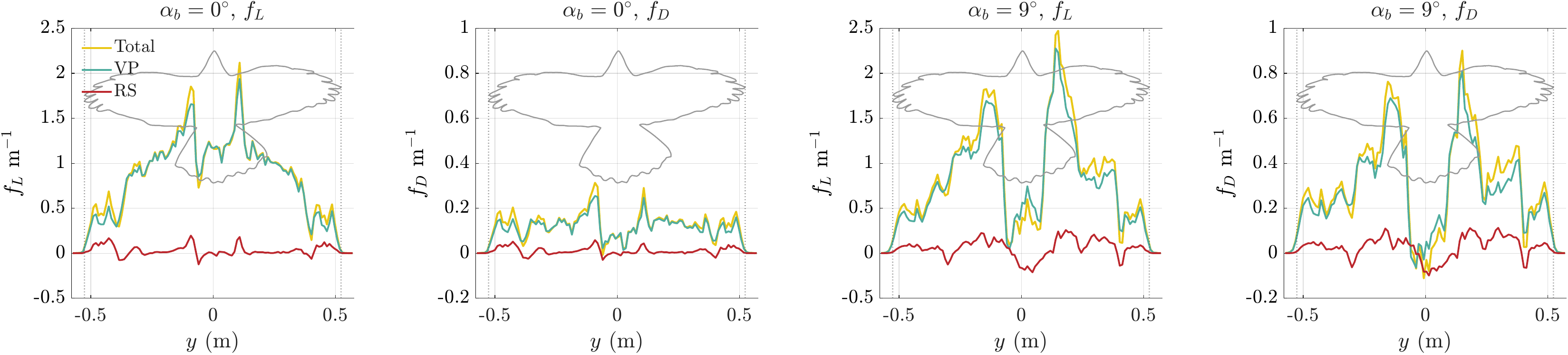}
\caption{Isosurfaces of the VP (a,c) RS (b,d) lift contribution with streamlines around the gliding bird ($1^{st}$ row) . Spanwise distribution of lift (a,c) and drag (b,d) ($2^{nd}$ row). (a,b) $\alpha_b = 0^{\circ}$and(c,d) $9^{\circ}$.}
\label{fig:placeholder}
\end{figure}
 
Figure~\ref{fig:Results} shows incidence-matched flow fields and dominant lift components for the aerofoil and the bird's mid-span wing section. In the first and second rows, the VP contribution $\fvpp_{L} \!\approx\! \fvpp_{L,y}$ arises from the combined effect of $\omega_y$ and $\Lambda_{L,y}$. For the aerofoil and the bird wing section, flow fields at small $\alpha$ are similar, while they differ markedly at high $\alpha$ owing to 3-D effects. At small $\alpha$, the spanwise vorticity above the upper surface produces a positive lift contribution, while the wake region contributes negatively; the opposite pattern occurs on the lower surface and in the corresponding part of the wake. At large $\alpha$, the aerofoil exhibits alternating shedding of the leading-edge vortex (LEV) and trailing-edge vortex (TEV); a representative snapshot in one cycle is shown. The upper portion of the LEV and the lower portion of the TEV contribute positively to lift, while the opposite regions contribute negatively. In contrast, the bird wing section maintains a stable vortex structure due to the 3-D effect, with both LEV and TEV generating positive lift of much greater magnitude than in the aerofoil. Nevertheless, the overall lift of the bird is expected to be lower when averaged over the entire body and tail, as explained later.
The lift components in other directions are shown in Appendix~\ref{app:figs} (Figure~\ref{fig:app_vp_fields}).

The third and fourth rows of Figure~\ref{fig:Results}, diagnose the RS contribution from $\bm{\zeta}$ using contour plots $\phi_{L~z}$. At low $\alpha$, the aerofoil's largely 2-D, attached flow produces similar $\zeta_z$ concentrations on the upper and lower surfaces and its wake; when weighted by $\Bnabla\phi_L$, they lead to equal and opposite net lift force yielding a negligible net $\Frs_{L}$. In contrast, the bird's section exhibits inherently 3-D features leading to asymmetrical $\zeta_z$ values, with stronger regions on the upper surface and in the trailing edge (TE) wake shed from the lower surface remaining aligned with $\Bnabla\phi_L$; this disparity leads to a net positive $\Frs_{L}$ appreciable even at low $\alpha$. At higher $\alpha$, both geometries are strongly separated, so RS lift grows overall, with a particularly marked rise for the aerofoil once alternating shedding is established and breaks the symmetry. For the bird, sustained 3-D roll-up keeps $\zeta_z$ aligned with $\Bnabla\phi_L$ over broad regions, and RS lift remains substantial across all $\alpha$ (Figures~\ref{fig:Results_plots} and \ref{fig:Results}). The corresponding $x$- and $y$-components of the RS lift density are shown in Appendix~\ref{app:figs} (Figure~\ref{fig:app_rs_fields}).

Figure~\ref{fig:placeholder} provides a complementary 3-D view of the VP and RS lift contributions for the bird at $\alpha_b = 0^\circ$ and $9^\circ$. At $\alpha_b = 0^\circ$, the lift-generating regions remain confined to a thin layer enveloping the body and extending into the wake. Both VP and RS contributions exhibit similar patterns, with positive lift concentrated along the upper surface and in the lower-surface TE wake. The rolling-up wingtip vortex (WTV) also produces a pronounced positive lift contribution. At $\alpha_b = 9^\circ$, large-scale separation dominates the lift generation over the wing sections, with substantial contributions arising from the LEV, TEV, and the wingtip region where the rolling-up WTV remains a strong source of positive lift. In contrast, the body section contributes comparatively little: the flow is fully separated, and the positive and negative lift-producing regions largely cancel, as further demonstrated in appendix~\ref{app:figs}. 
Moreover, from Figure~\ref{fig:Results} and Figure~\ref{fig:placeholder}, we can see that as both $|\Lambda_{L,y}|$ and $\phi_{k,z}$ (or, more generally, $\Bnabla\phi_k$) decay rapidly away from the body, the VP and RS force contributions are concentrated within roughly two chord lengths of the surface. An analogous decomposition holds for drag: $\fvpp_{D}$ is dominated by $\omega_y$ at low~$\alpha$ for the aerofoil, while for the bird $\frsp_{D}$ becomes significant as $\zeta_z$ aligns with $\Bnabla\phi_D$ at higher incidence, which are not shown here.

\section{Conclusion}\label{sec:conclusion}
The RA-VFM has been derived by including an RS contribution, which is the scalar product of $\Bnabla\phi_k$ and the divergence of the Reynolds stress ($\zRS=\Bdiv\taurs$) estimated from the Boussinesq approximation. Applied to a goshawk in a gliding configuration and its incidence-matched aerofoil, RA-VFM reproduces the CFD force curves while retaining the diagnostic value of the original VFM method. For the aerofoil, the VP term $\Fvp$ alone is sufficient except for post-stall conditions; for the bird, $\Frs$ is essential due to 3-D effects. 
As $\Bnabla\phi_k$ decays rapidly away from the body, only the near-body flow contributes appreciably to the force integral. Consequently, accurate force evaluation and flow-structure attribution can be achieved using a compact domain surrounding the object. 
The approach is therefore directly applicable to turbulent flows whenever Reynolds stresses are available from closure modelling or measurements, and provides a practical route to linking mean aerodynamic loads to specific flow structures in complex three-dimensional geometries.

\section*{Declarations and Acknowledgements}
Acknowledgements:

This research is funded by Engineering Start-up Grant of King's College London, and the Daiwa Anglo-Japanese Foundation through Daiwa Foundation Awards (14465/15310). We thank Prof. Richard Bomphrey and Dr. John Cheney for the experimental data.

Declaration of interests:

The authors report no conflict of interest.

AI-assisted tools:

AI-assisted tools were used during manuscript preparation for grammar and language polishing, in accordance with the journal’s policy and requirements.

\bibliographystyle{plainnat}
\bibliography{references}
\newpage
\appendix
\section{Component-wise form of the vortex-pressure term}\label{app:vp_split}
Here we derive the component-wise form of~\ref{eq:RAVFM_vp}
\begin{equation}\label{eq:app_vp_start}
\Fvp_{k}
=\rho \iiint_{\Omega} \Bnabla\phi_k \cdot \bigl(\Um \times \wmean\bigr)\dO.
\end{equation}
Using the anti-symmetry property of the scalar triple product
\begin{equation}\label{eq:app_vp_triple}
\Bnabla\phi_k \cdot \bigl(\Um \times \wmean\bigr)
= \wmean \cdot \bigl(\Bnabla\phi_k \times \Um\bigr),
\end{equation}
and substituting $\Um=(u,v,w)$, $\wmean=(\omega_x,\omega_y,\omega_z)$, 
$\Bnabla\phi_k=(\phi_{k,x},\phi_{k,y},\phi_{k,z})$
into\eqref{eq:app_vp_start} yields 
\begin{equation}\label{eq:app_vp_componentwise}
\Fvp_{k}
=\rho \iiint_{\Omega}\!\Big[
\omega_x\bigl(\phi_{k,y}w-\phi_{k,z}v\bigr)
+\omega_y\bigl(\phi_{k,z}u-\phi_{k,x}w\bigr)
+\omega_z\bigl(\phi_{k,x}v-\phi_{k,y}u\bigr)
\Big]\dO.\
\end{equation}

An equivalent form of~\ref{eq:app_vp_componentwise} is 
\begin{equation}\label{eq:app_vp_Lambda}
\Fvp_{k}
=\rho \iiint_{\Omega}\!\big[
\omega_x\,\Lambda_{k,x}\!\cdot(0,v,w)
+\omega_y\,\Lambda_{k,y}\!\cdot(u,0,w)
+\omega_z\,\Lambda_{k,z}\!\cdot(u,v,0)
\big]\dO,\
\end{equation}
where the geometry-based vectors
\begin{equation}\label{eq:app_Lambda_def}
\Lambda_{k,x}=(0,-\phi_{k,z},\ \phi_{k,y}),\qquad
\Lambda_{k,y}=(\phi_{k,z},0,-\phi_{k,x}),\qquad
\Lambda_{k,z}=(-\phi_{k,y},\ \phi_{k,x},0).
\end{equation}
\newpage

\section{Additional force components analysis}\label{app:figs}
\begin{figure}
\centering
\includegraphics[width=\linewidth,trim=0 2.1cm 0 2cm,clip]{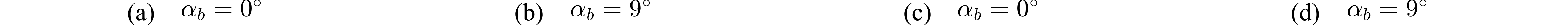}
\includegraphics[width=1\linewidth]{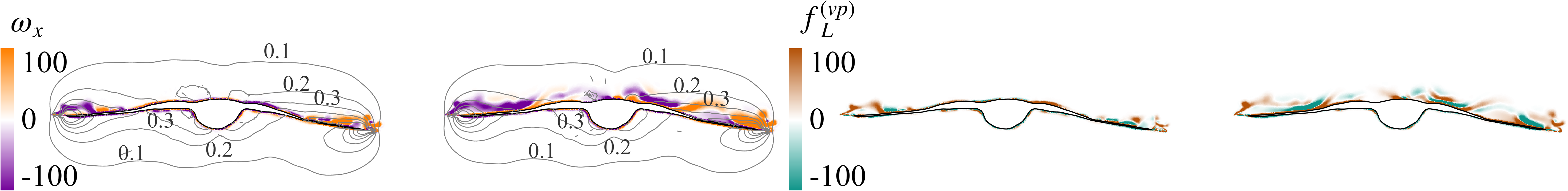}
\includegraphics[width=\linewidth,trim=0 2.1cm 0 2cm,clip]{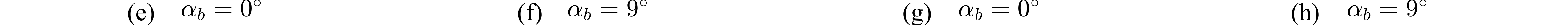}
\includegraphics[width=1\linewidth]{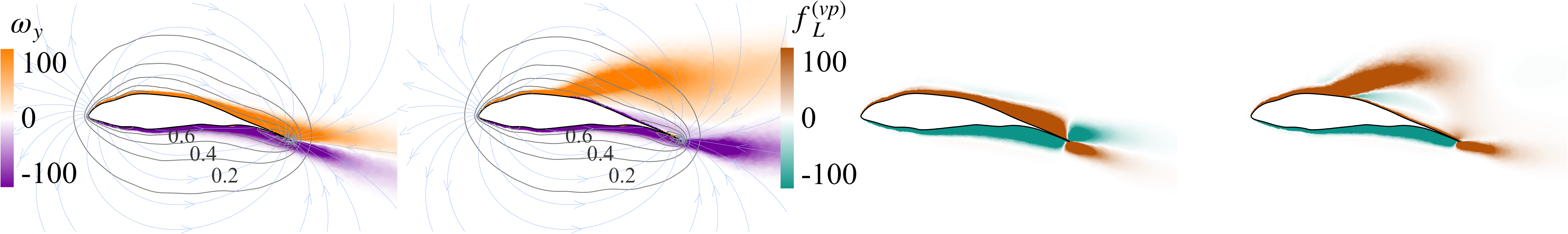}
 \caption{Visualisation of vorticity fields and VP lift for the bird at $\alpha_b = 0^{\circ}$ and $9^{\circ}$. (a–d) $x$-normal sections: (a,b) contours of $\omega_x$ overlaid with $|\Lambda_{L,x}|$ isolines, and (c,d) the corresponding VP lift contribution $f^{(\mathrm{vp})}_{L}$. (e–h) The body section ( $y=0$). 
} \label{fig:app_vp_fields}
\end{figure}

Figure~\ref{fig:app_vp_fields} shows section views of the VP lift components for the bird. 
Overall, the VP lift contribution in the $x$-direction is weak: at low $\alpha_b$ the wingtip vortices provide positive contributions primarily, while at higher incidence the pattern becomes more fragmented and chaotic as separation develops. The strong separation at the centre-body section at higher incidence results in smaller VP force contributions compared to the small incidence case, which further supports the conclusion from Figure~\ref{fig:placeholder}. 

\begin{figure}
\centering
\includegraphics[width=1\linewidth]{Figs/Results/fig_fields_header.pdf}
\includegraphics[width=1\linewidth]{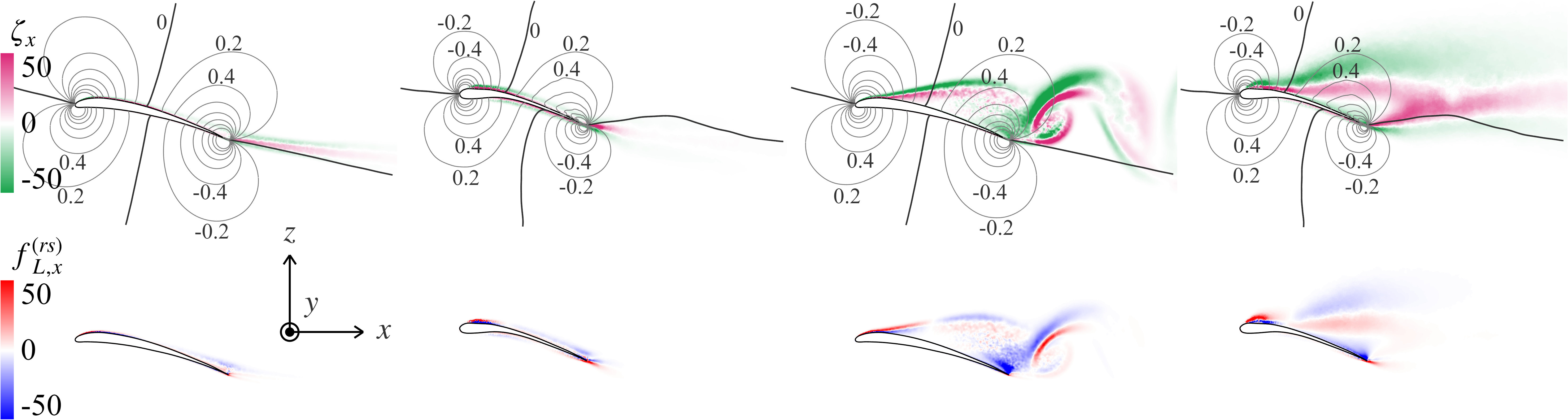}
\includegraphics[width=1\linewidth]{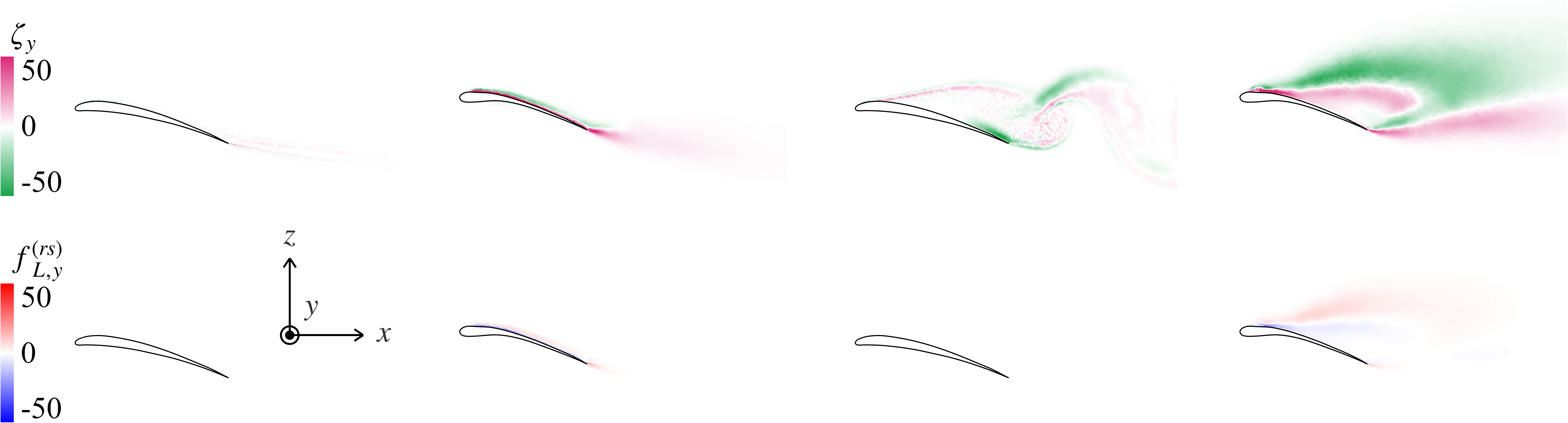}
\caption{Non-dominant RS lift components at incidence-matched conditions ($\alpha_{\mathrm{eff}} = \alpha = 13^{\circ}$ and $22^{\circ}$) for the aerofoil (a,c) and bird (b,d). Row~1 shows contours of $\zeta_x$ overlaid with $\phi_{L,x}$ isolines, and row~2 shows the corresponding RS lift contribution $f^{(\mathrm{rs})}_{L,x} = \phi_{L,x}\,\zeta_x$. Row~3 shows contours of $\zeta_y$ overlaid with $\phi_{L,y}$ isolines, and row~4 the RS lift density $f^{(\mathrm{rs})}_{L,y} = \phi_{L,y}\,\zeta_y$.
}
\label{fig:app_rs_fields}
\end{figure}

Figure~\ref{fig:app_rs_fields} shows the non-dominant RS lift components for the aerofoil and the bird's sections in the $x$- and $y$-components.
The $x$-component yields a small net \emph{negative} contribution to $\Frs_L$, concentrated near the upper trailing edge and strengthening at higher incidence. For the $y$-component, $|\zeta_y|$ is relatively strong for both aerofoil and bird, especially at higher $\alpha$. However, a zero $\phi_{L,y}$ for the aerofoil and a near-zero $\phi_{L,y}$ at the mid-span bird section results in a weak net RS lift contribution.

\end{document}